\documentclass[useAMS,usenatbib]{mnras}

\usepackage{graphicx}
\usepackage{amssymb}

\newcommand{\kms}{km\,s$^{-1}$}
\newcommand{\kmsmpc}{km\,s$^{-1}$\,Mpc$^{-1}$}
\newcommand{\Msun}{\mbox{$M_{\sun}$}}
\newcommand{\MsunLsun}{$M_{\sun}/L_{\sun}$}

\title[An isolated pair of nearby dwarfs]
{DDO 161 and UGCA 319: an isolated pair of nearby dwarf galaxies}

\author[
I.\ D.\ Karachentsev,
L.\ N.\ Makarova, 
R.\ B.\ Tully, 
L.\ Rizzi, 
V.\ E.\ Karachentseva
\& 
E.\ Shaya
]
{I.\ D.\ Karachentsev,$^1$\thanks{E-mail: ikar@sao.ru}
L.\ N.\ Makarova,$^1$
R.\ B.\ Tully,$^2$
L.\ Rizzi,$^3$
V.\ E.\ Karachentseva$^4$
\newauthor
and
E.\ J.\ Shaya$^5$
\\
$^1$ Special Astrophysical Observatory of RAS, Nizhnij Arkhyz, KChR, 369167, Russia\\
$^2$ Institute for Astronomy, University of Hawaii, 2680 Woodlawn Drive, HI 96822, USA\\
$^3$ W. M. Keck Observatory, 65-1120 Mamalahoa Hwy, Kamuela, HI 96743, USA\\
$^4$ Main Astronomical Observatory, Academy of Sciences of Ukraine, 27 Akademika Zabolotnoho St, UA-03143 Kyiv, Ukraine\\
$^5$ Astronomy Department, University of Maryland, College Park, MD 20743, USA
}

\begin{document}

\label{firstpage}

\pagerange{\pageref{firstpage}--\pageref{lastpage}}

\pubyear{2016}

\maketitle

\begin{abstract}

We report \textit{HST}/ACS observations of two nearby gas-rich dwarf galaxies: 
DDO\,161 and UGCA\,319. Their distances determined via the Tip of the Red
Giant Branch are $6.03_{-0.21}^{+0.29}$\,Mpc and $5.75\pm0.18$\,Mpc,
respectively. The galaxies form an isolated pair dynamically well separated
from the nearest neighbors: KK\,176 ($7.28\pm0.29$\,Mpc) and NGC 5068 ($5.16\pm0.21$\,Mpc). 
All four galaxies have a bulk spatial peculiar velocity towards the
Virgo cluster of $\sim158\pm17$\,\kms{} in the Local Group rest frame and $\sim330$\,\kms{}
with respect to the cluster center.

\end{abstract}


\section{Introduction}

\citet{tully2002} drew attention to the existence of a special category 
of systems of galaxies containing only galaxies of low luminosity, like 
Magellanic clouds or below. Subsequent accurate measurements of their
distances with the Hubble Space Telescope (\textit{HST}) using the Tip of the Red 
Giant Branch (TRGB) confirmed that dwarf galaxies in these nests are
associated with each other not only on the sky but also by the spatial
distances \citep{tully2006}. Their closest example is
a quartet of dwarfs: NGC\,3109, Sex\,A, Sex\,B and Antlia, located just beyond 
the Local Group threshold at a distance of $\sim1.4$\,Mpc. Later, the quartet
was transformed into elongated sextet due to the detection of two ultra-dwarf
galaxies: Antlia\,B and Leo\,P \citep{mcquinn2013,sand2013}. 
According to \citet{tully2006}, associations of dwarfs are characterized by
a dimension of $\sim400$\,kpc, a radial velocity dispersion of $\sim30$\,\kms{}, a total
stellar mass of $\sim10^9$\,\Msun{} and a virial-to-stellar mass ratio of $\sim(300 - 1000)$.
Because of low luminosity, associations of dwarfs are detectable predominantly
in the nearby volume.

Considering a sample of about 10000 galaxies in the nearby universe with
radial velocities $V_{LG} < 3500$\,\kms{} respect to the Local Group rest frame, 
\citet{kar2008} and \citet{makarov2011}
clusterized this population into groups of different multiplicity. 
Their clustering algorithm took into account the large observed diversity of galaxies 
in luminosities. In this sample, \citet{kar2008}
noted the unexpected abundance of binary dwarf galaxies. \citet{makarov2012} 
compiled a list 57 systems in the volume of $V_{LG} < 3500$\,\kms{} 
formed by two or more dwarf galaxies. A typical size of these MU-groups is
$\sim30$\,kpc, a typical velocity dispersion is only $\sim10$\,\kms{}. At the mean
stellar mass of group of dwarfs about $3\times10^8$\,\Msun{}, their typical 
virial-to-stellar mass ratio turns out to be $\sim40$. The authors noted 
a lack of gap in size and luminosity between the MU-groups and associations
of dwarfs discussed by \citet{tully2006}. However, while a "crossing time" for 
associations of dwarfs is $\sim9$\,Gyr, i.e. comparable with the age of the universe,
the mean crossing time for MU-groups is only $\sim3$\,Gyr, being typical for 
dynamically relaxed systems. 

As has been noted by \citet{makarov2012}, pairs and groups of dwarfs
prefer to reside in regions of low spatial density, and their components
are usually gas-rich dwarfs. Observing the population of nearby voids in the
21 cm HI line with the Giant Metrewave Radio Telescope, \citet{cheng2013} and 
\citet{cheng2017} found several multiple gas-rich dwarfs of
very low metallicity.  Presumably, a thorough HI-survey of nearby dwarfs with
the spatial resolution offered by interferometers could  significantly extend the 
existing collection of multiple dwarf galaxies.

In this Note, we report on measurements of accurate distances to dwarf
galaxies DDO\,161 = UGCA\,320 and UGCA\,319 with radial velocities 543\,\kms{} and
555\,\kms{} respectively, which form a tight pair well separated from other
neighboring galaxies.

\section{\textit{HST} observations and data reduction}

\begin{figure*}
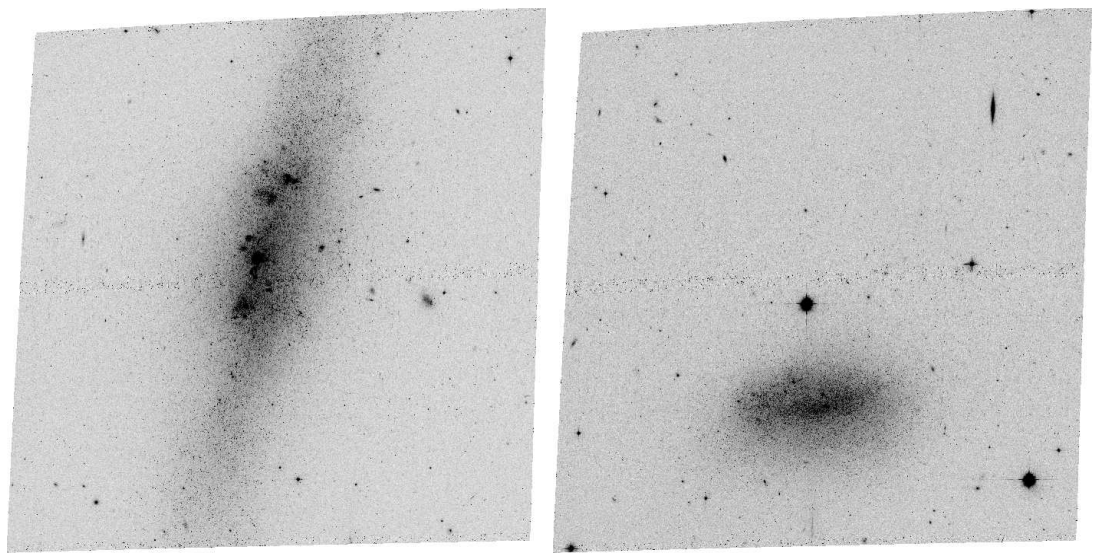

\includegraphics[width=0.4\textwidth,clip]{fig1_1.ps}
\includegraphics[width=0.4\textwidth,clip]{fig1_2.ps}
\caption{
\textit{HST}/ACS images of DDO\,161 (left panel) and UGCA\,319 (right panel) in the 
\textit{F606W} filter. The dimension of the fields is about 3.3 arcmin. North is
straight out to the upper left corners of the images, and East is to the lower left ones.}
\end{figure*}

The galaxies DDO\,161 and UGCA\,319 were observed with the
Hubble Space Telescope (\textit{HST}) using Advanced Camera for Surveys (ACS)
on February 11 and 14, 2017 (cycle 24 proposal GO-14636, PI Karachentsev).
Four exposures were made in a single orbit per galaxy: \textit{F606W} (broad-
band $V$), \textit{F814W} (broad-band $I$) filters at the same position as well as
\textit{F606W} and \textit{F814W} at a dithered position, about 3 arcsec away.
The total exposure time per galaxy for each filter was 1030 seconds.
The \textit{F606W} images of the galaxies are shown in Fig.\,1. 
The dimension of the fields is about 3.3 arcmin. 

The photometry of
resolved stars in the galaxies were performed with the ACS module of the
DOLPHOT package (http://purcell.as.arizona.edu/dolphot/) for crowded
field photometry \citep{dolphin2002} using the recommended recipe and
parameters. Only stars with photometry of good quality were included
in the final compilation, following recommendations given in the DOLPHOT
User's Guide. Artificial stars were inserted and recovered using the
same reduction procedures to accurately estimate photometric
errors, including crowding and blending effects. A large library of
artificial stars was generated spanning the full range of observed
stellar magnitudes and colours to assure that the distribution of the
recovered photometry is adequately sampled.

\begin{figure*}
\includegraphics[width=0.3\textwidth,clip]{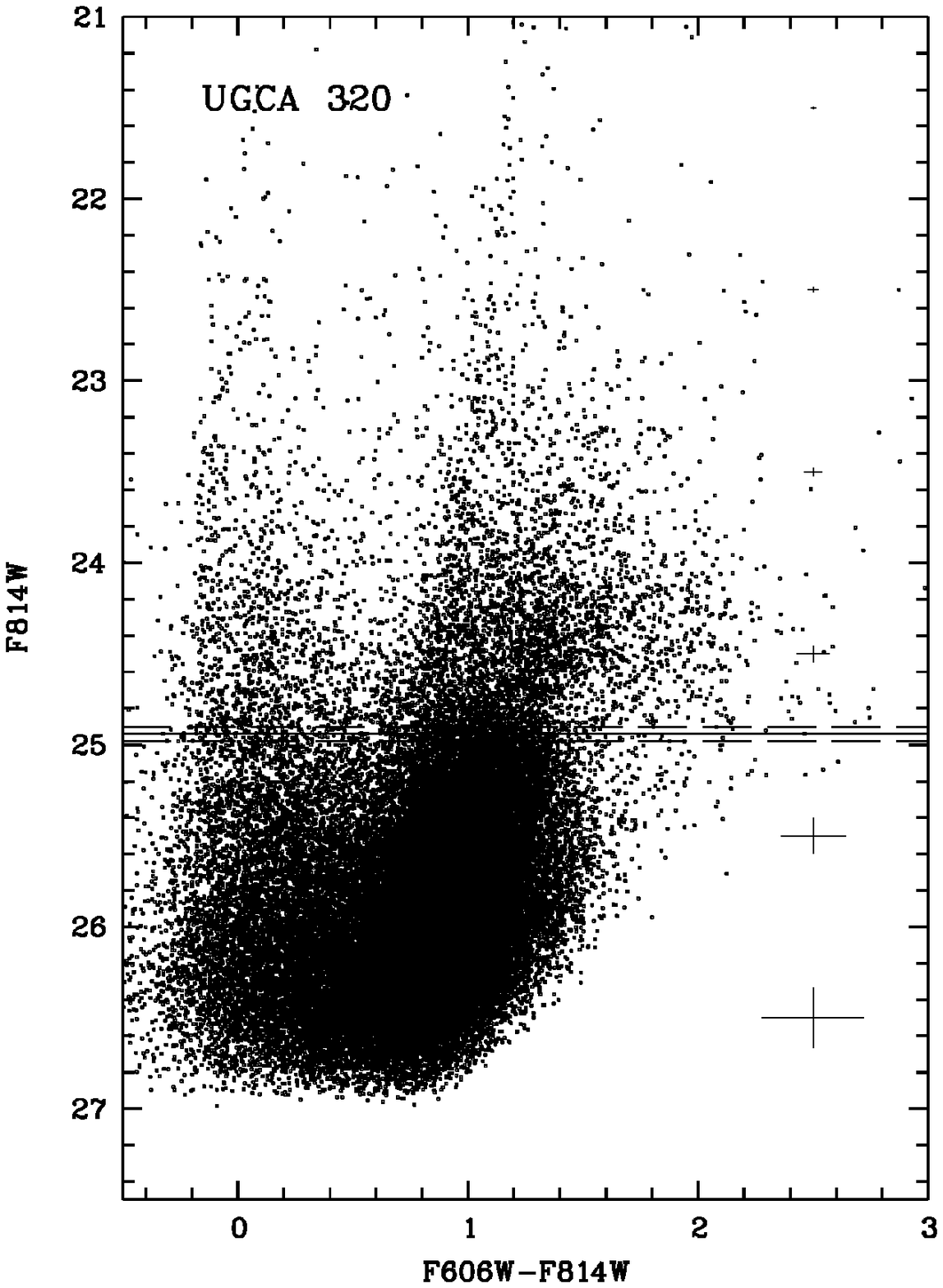}
\includegraphics[width=0.3\textwidth,clip]{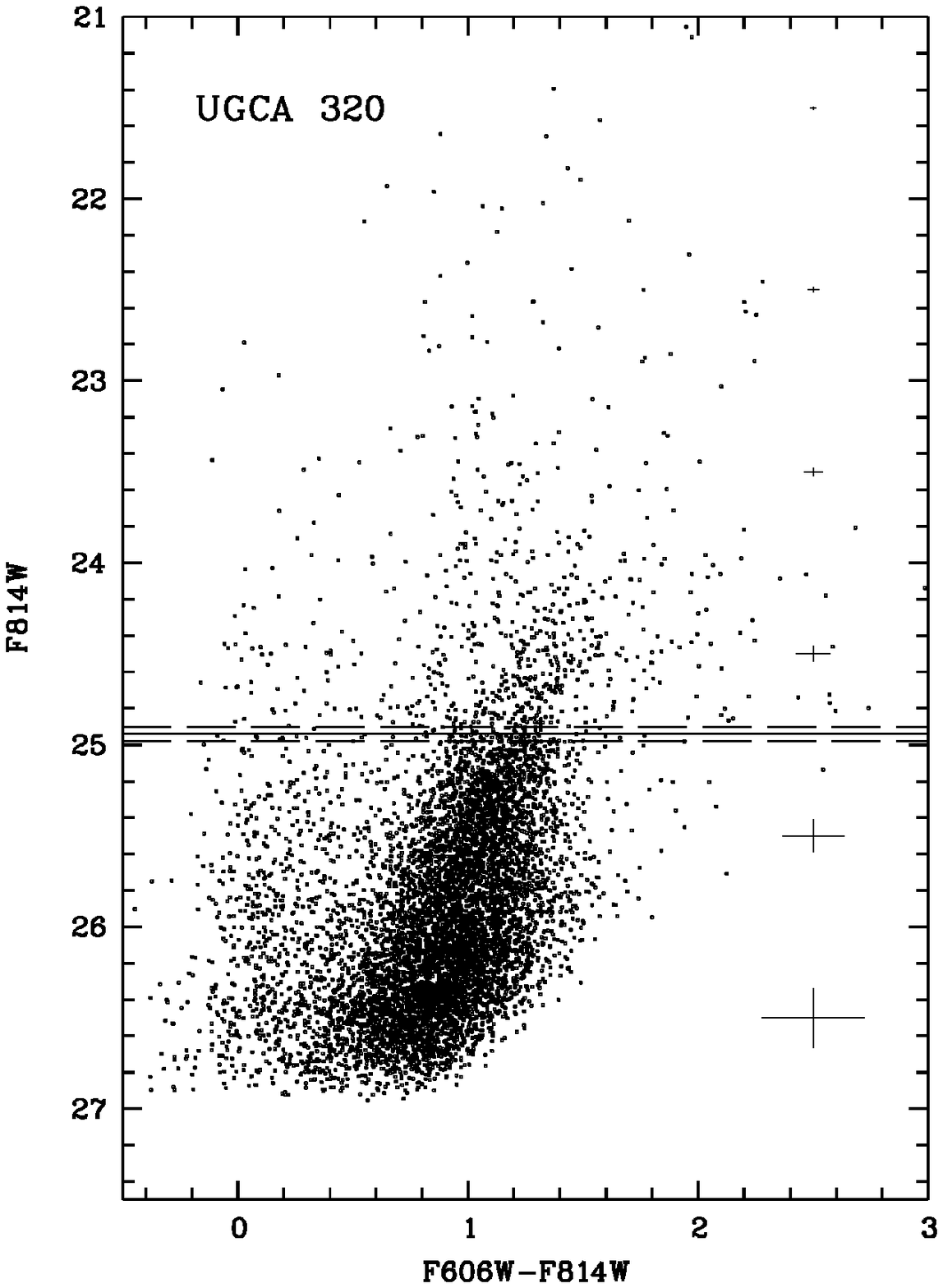}
\includegraphics[width=0.3\textwidth,clip]{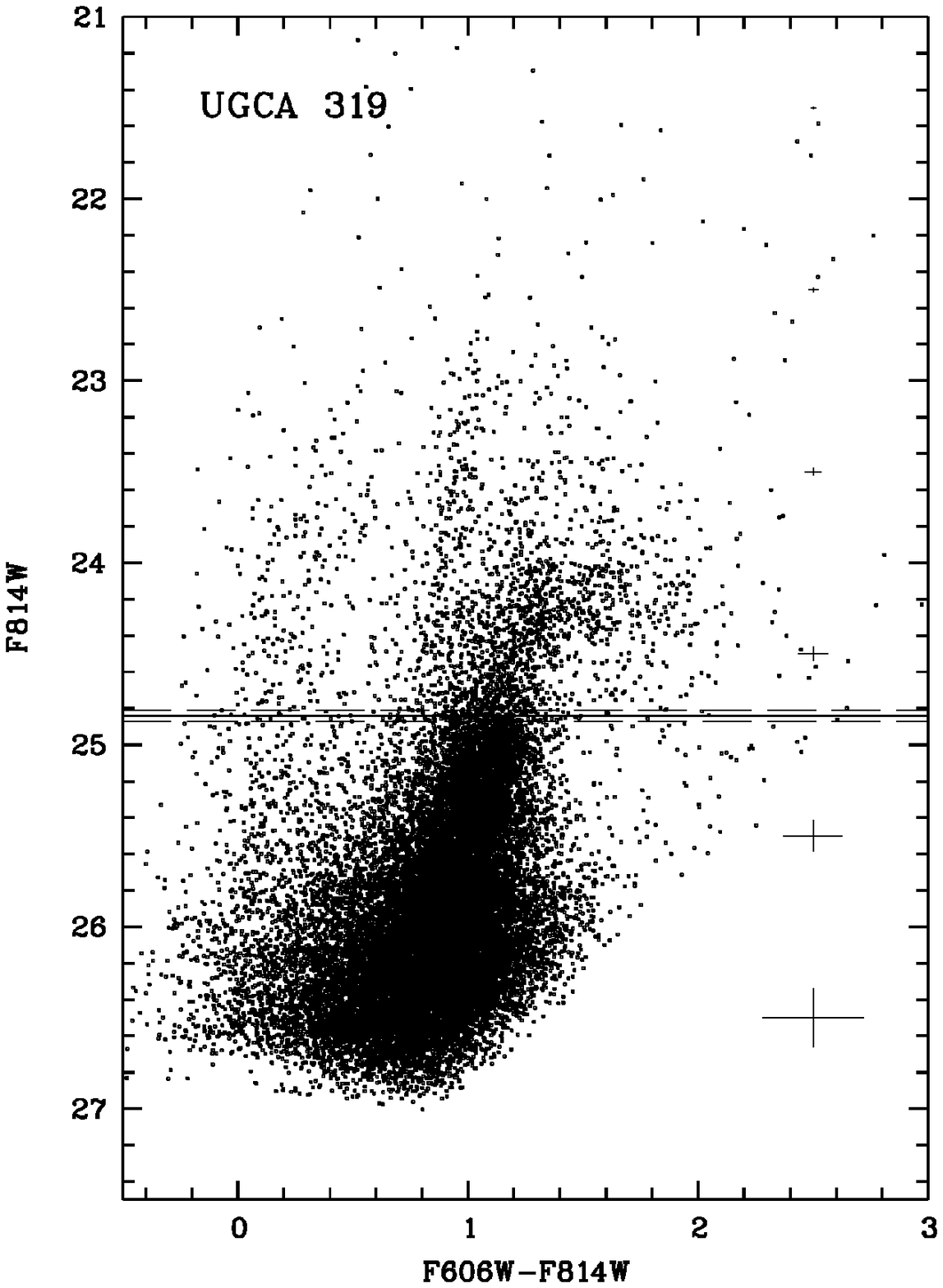}
\caption{
DDO\,161 = UGCA\,320 and UGCA\,319 colour-magnitude diagrams. 
Photometric errors are indicated by the bars at the right in the CMD. }
\end{figure*}

Colour-magnitude diagrams (CMDs) for the detected stars in DDO\,161 and
UGCA\,319 are shown in Figure\,2. In the case of DDO\,161 we present the CMD
for the complete galaxy body (left panel) and after excluding some
central regions crowded by star formation complexes (middle panel).
The method of deriving the distance from the Tip of the Red Giant Branch
(TRGB) magnitude rests on a well understood physical basis \citep{madore1997}.
Observationally, there is a sharp cutoff of the bright end of
the red giant branch luminosity function at $M_I\simeq-4.0$\,mag.

\begin{table}
\caption{TRGB and distance measurements for the pair components.}
\begin{tabular}{lll} \hline

 TRGB parameters  &       DDO161      &    UGCA319\\
\hline
 $F814W$              &  $24.94_{-0.04}^{+0.08}$  &  $24.84\pm0.03$\\

$\langle F606W-F814W\rangle$   &    $1.07\pm0.01$   &  $1.01\pm0.01$\\

 $E(B-V)$              &    $0.070$     &      $0.072$ \\

$(m-M)_0$             &   $28.90_{-0.07}^{+0.10}$   & $28.80\pm0.07$\\

 D, Mpc               &    $6.03_{-0.21}^{+0.29}$  &   $5.75\pm0.18$\\ \hline
\end{tabular}
\end{table}

We determined the TRGB using a maximum-likelihood algorithm
written and provided by \citet{makarov2006}. The estimated values of
the TRGB and their errors are presented in Table\,1. The measured TRGB
positions are marked in Fig.\,2 by the solid line. The absolute magnitude
of the TRGB in the \textit{HST} flight system for F606W and F814W using ACS
was estimated using the calibration from \citet{rizzi2007}:
$$M(F814W) = - 4.06 + 0.20 [(F606W - F814W) - 1.23].$$ 
The formula defines a zero-point calibration of the TRGB as a function of the stellar
population colour, accounting for variation due to metallicity
and age. As a result, we derived the distances: $6.03_{-0.21}^{+0.29}$ Mpc
and $5.75\pm0.18$ Mpc for DDO\,161 and UGCA\,319 respectively. These
measurements assume foreground Galactic extinction, $E(B-V)$,
according to \citet{extin2011}. Based on the accurate TRGB distance 
determinations, we conclude that DDO\,161 and UGCA\,319 are spatially associated 
each other forming a physical system.   The measured distances differ by $280\pm310$~kpc.

\section{Discussion}

NASA Extragalactic Database (=NED, http://ned.ipac.caltech.edu) contains
10 distance estimates for DDO\,161 made by the Tully-Fisher \citep{tully1977} method
in its different modifications. The average value among them, 6.1\,Mpc, is
perfectly consistent with the TRGB distance 6.03 Mpc measured by us. In contrast to 
DDO\,161, NED presents for UGCA\,319 only a single distance estimate, 19.7\,Mpc,
via the infra-red Tully-Fisher relation \citep{sorce2014} that is in sharp
contradiction to our present TRGB measurement. The nature of this discrepancy
is caused apparently by confusion of the HI-line width of UGCA\,319 with the 
strong HI-flux from the close neighbor DDO\,161. This region of sky has been imaged
in HI with Australian Telescope Compact Array (ATCA) by \citet{pisano2011}.
Answering our request, Daniel Pisano kindly re-estimated HI parameters for UGCA\,319 
from ATCA observations.

\begin{table}
\caption{Global properties of the pair components}
\begin{tabular}{lcc} \hline

 Parameter                   &       DDO 161     &    UGCA 319\\
\hline
R.A.,Dec.(J2000.0)            &  130316.8--172523 & 130214.4--171415\\
Morphological type             &        Sm      &        Ir\\
$B_T$, mag                      &       13.5    &       15.1 \\
Holmberg diameter, arcmin     &        7.94   &         1.26\\
$V_{LG}$,  \kms{}                   &        543    &        555 (529)\\
$W_{50}$,  \kms{}                   &        113    &         94 (30)\\
$\log(F_{HI})$, Jy \kms{}            &         1.88  &          0.88 (-0.20)\\
$m(FUV)$                        &        14.6   &         17.2 \\
Distance,  Mpc                &         6.03  &          5.75\\
$M_B$,  mag                     &        --16.04  &        --14.17\\
$\log(M_*/$\Msun{}$)$                  &        8.75   &         8.01\\
$\log(M_{HI}$/\Msun{}$)$                &        8.83   &         7.76 (6.68)\\
$\log(sSFR)_{FUV}$, yr$^{-1}$         &      --9.80   &       --10.30 \\
\hline
Projected separation, arcmin        &  \multicolumn{2}{c}{18.6}\\
Projected separation, kpc           &     \multicolumn{2}{c}{32.7}\\
Orbital mass,$10^9$\,\Msun{}             &  \multicolumn{2}{c}{5.6 (7.6)}\\
$M_{orb}/(M^*_1 + M^*_2)$                 & \multicolumn{2}{c}{8.3 (11.3)}\\
$T_{cross}$, Gyr                        &   \multicolumn{2}{c}{2.7 (2.3)} \\ \hline
\end{tabular}
\end{table}  

The basic properties of DDO\,161 and UGCA\,319 taken from the NED and HyperLEDA
\citep{makarov2014} databases are presented in Table\,2. The data for UGCA\,319
from D.Pisano are given in parentheses. The table contains:
equatorial coordinates; morphological type; total B-magnitude; Holmberg's
diameter; radial velocity; 21\,cm line width at the 50\%-level; 
the HI line flux; apparent magnitude
in the Far Ultra-Violet \citep{gil2007}; distance measured via TRGB;
absolute magnitude corrected for Galactic and internal extinction; stellar mass
derived via the K-band luminosity from the Local Volume Galaxy Data Base (http://www.sao.ru/lv/lvgdb/)
assuming $M_*/L_K = 1\times$\MsunLsun{}; hydrogen mass and
specific star formation rate 

$$log(sSFR) = 2.78 - 0.4\times m(FUV) + 2\times log(D_{Mpc}) - log(M_*/\Msun{})$$ 

calculated via the FUV-flux \citep{lee2011}. 

As one can see, the galaxies are gas-rich dwarf systems. Their present
star formation rate is moderate, but sufficient to reproduce the observed stellar 
mass of the galaxies over the age of universe, 13.7\,Gyr.

DDO\,161 and UGCA\,319 as a physical pair of dwarfs are present in the lists by
\citet{kar2008} and \citet{makarov2012}. The linear projected 
separation of the components, $R_p$, is 32.7\,kpc. At the radial velocity difference
$\Delta V$ = 12 (14)\,\kms{} the orbital mass of the pair,

$$M_{orb} = (16/\pi \times G)\times R_p\times\Delta V^2$$ 

is $5.6\, (7.6)\times10^9$\,\Msun{}, being 8.3 (11.3) times the sum
of stellar masses of the components. These parameters of the binary dwarf system
are presented in the last rows of Table\,2 along with the crossing time,
$t_{cross} = R_p / \Delta V \sim 2.7\, (2.3)$\,Gyr.

Note that the first estimates of dynamical mass of DDO 161 have been derived
by \citet{fisher1975} and \citet{kar1981} from
the HI line width $W = 2\times V_m = 114\pm7$ \kms{} and from the optical rotation amplitude of 
$V_m \simeq 50$\,\kms{}, respectively.
The implied internal dynamical mass, $\sim(1.5 - 4.7)\times10^9$\,\Msun{}, is in a reasonable agreement with 
the orbital mass estimate, $(5.6 - 7.6)\times10^9$\,\Msun{}, presented in this Note. 

\section{The pair environment}

\begin{figure*}
\includegraphics[width=0.3\textwidth,clip]{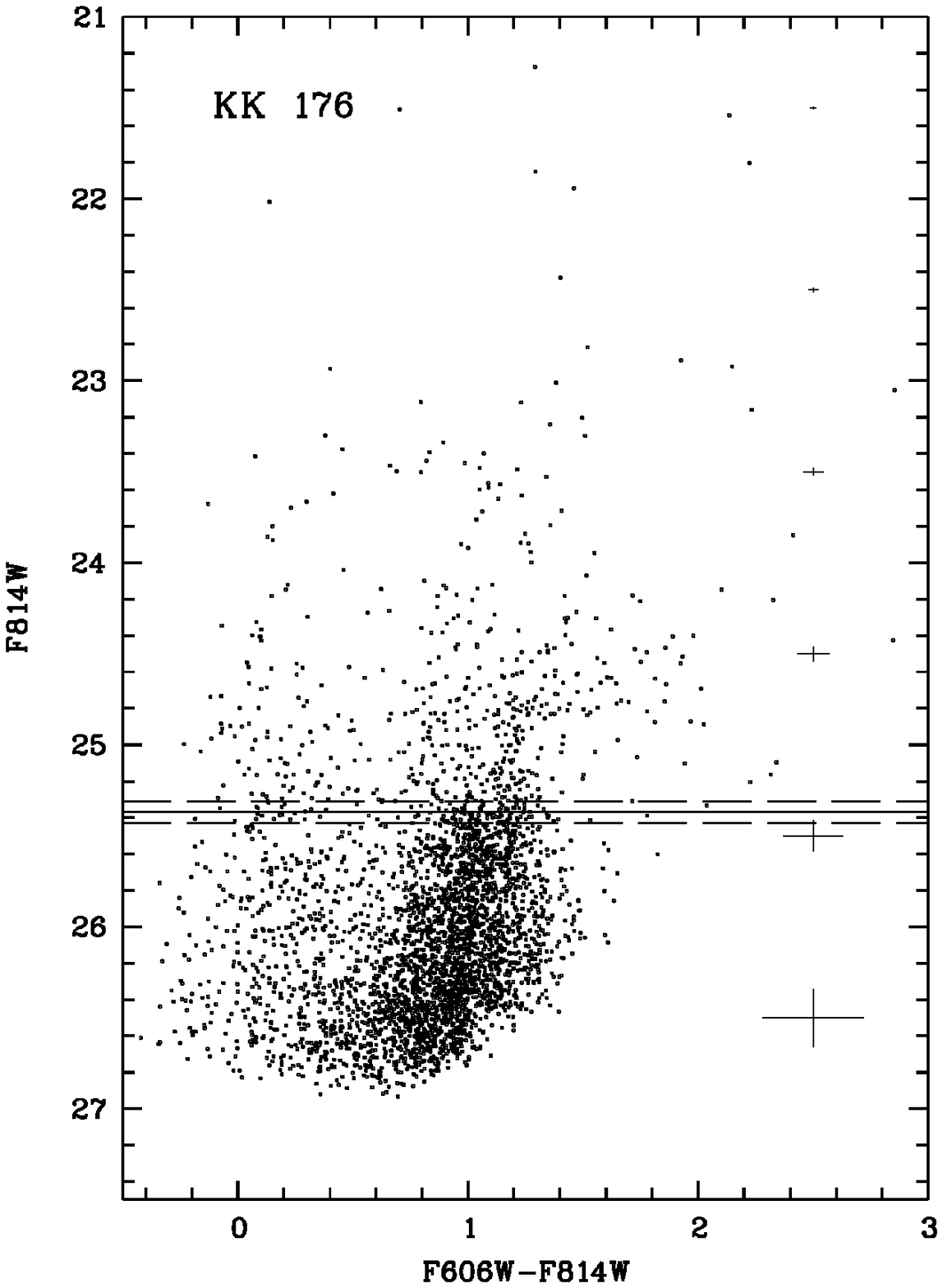}
\includegraphics[width=0.3\textwidth,clip]{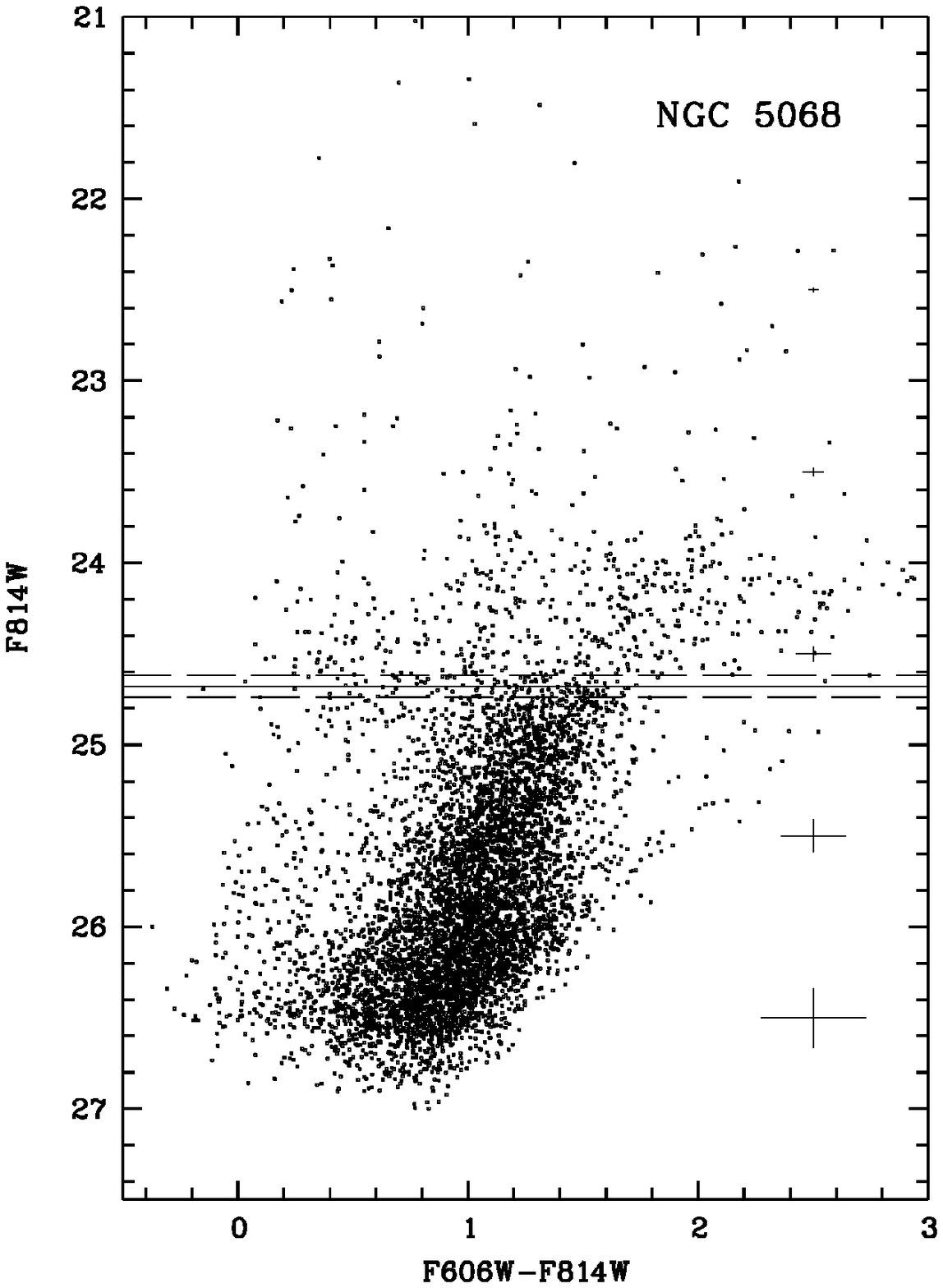}
\caption{
KK\,176 and NGC\,5068 colour-magnitude diagrams. 
Photometric errors are indicated by the bars at the right in the CMD. }
\end{figure*}

Apart from UGCA\,319, in the wide surroundings of DDO\,161 up to the angular 
separation of $\sim5^o$ there are only two galaxies with radial velocities within
$\pm200$\,\kms{} around the velocity of DDO\,161: the dwarf galaxy KK\,176 at a separation
of 128 arcmin with the velocity of $V_{LG} = 618$\,\kms{}, and the spiral galaxy NGC\,5068 at a
separation of 310 arcmin with the velocity of $V_{LG} = 469$\,\kms{}. \citet{pisano2011}
assumed that DDO\,161 together with KK\,176, NGC\,5068 and another dwarf BCD galaxy
MCG-03-34-002 ($V_{LG} = 765$\,\kms{}) form a loose "HIPASS"-group having the linear
diameter of $\sim600$\,kpc, radial velocity dispersion of $124$\,\kms{} and the virial mass of
$5.1\pm2.5\times10^{11}$\,\Msun{}. 

\begin{table}
\caption{TRGB and global parameters for KK\,176 and NGC\,5068}
\begin{tabular}{lcc} \hline
Parameter    &                    KK 176   &     NGC 5068 \\
\hline
$F814W$                        &  $25.37\pm0.06$  &  $24.68\pm0.06$ \\
$\langle F606W-F814W\rangle$   &  $1.08\pm0.01$   &  $1.31\pm0.03$  \\ 
$E(B-V)$                       &  $0.086$         &     $0.091$     \\
$(m - M)_0$                    & $29.31\pm0.09$   &     $28.56$     \\
 D, Mpc                        & $7.28\pm0.29$    &   $5.16\pm0.21$ \\ 
R.A.,Dec.(J2000.0)             & 125956.3--192447 & 131855.3--210221\\
Morphological type             &       Ir         &    Scd          \\
$B_T$, mag                     &      $16.5$      &      $10.5$     \\
Holmberg diameter, arcmin      &      $1.48$      &      $10.0$     \\
$V_{LG}$, \kms{}               &      $ 618$      &      $469$      \\ 
$W_{50}$,  \kms{}              &      $  37$      &       $74$      \\ 
$log(F_{HI})$, Jy \kms{}       &      $0.72$      &      $2.04$     \\  
$m(FUV)$                       &      $18.8$      &      $12.8$     \\
Distance,  Mpc                 &      $7.28$      &      $5.16$     \\
$M_B$,  mag                    &      $-12.97$    &     $-18.51$    \\
$log(M_*/\Msun{})$             &       $7.44$     &     $ 9.73$     \\
$log(M_{HI}/\Msun{})$          &       $7.81$     &      $8.82$     \\
$log(sSFR)_{FUV}$, yr$^{-1}$   &       $-10.34$   &      $-10.06$   \\ \hline
\end{tabular}
\end{table} 

We have observed the galaxies KK\,176 and NGC\,5068 with ACS on \textit{HST} during February 1
and February 11, 2017 in the same program with DDO\,161 and UGCA\,319. The CM diagrams
for both the galaxies are presented in Fig.\,3. The estimated values of the TRGB 
for KK\,176 and NGC\,5068 as well as basic properties of them are given in 
Table\,3. The distance estimates:
$7.28\pm0.29$\,Mpc for KK\,176 and $5.16\pm0.21$\,Mpc for NGC\,5068 
do not give grounds for considering these neighbors to be physically 
connected to the tight pair DDO\,161/UGCA\,319.  The larger galaxy NGC\,5068 is 540~kpc 
removed in projection and $730\pm310$~kpc removed in the line of sight.

It is interesting to note that all 4 galaxies with accurately measured distances
have positive peculiar velocities, $V_{pec} = V_{LG} - H_0\times D$ assuming the Hubble parameter
$H_0 = 73$\,\kmsmpc{}, i.e.: $+103$\,\kms{} (DDO\,161), $+135$\,\kms{} (UGCA\,319), 
$+87$\,\kms{} (KK\,176), 
and $+92$\,\kms{} (NGC\,5068). They are located in the Local supergalactic plane South from the
Virgo cluster center at angular separations of $(30 - 35)^o$. An apparent reason of 
their positive peculiar velocities is the infall of galaxies towards the Virgo cluster
as the nearest massive attractor. Assuming strictly radial infall towards Virgo,
we get the complete (spatial) vector of the mean peculiar velocity of the galaxy 
"flock" towards the cluster: $(104\pm11)/ cos(49^o) \simeq (158\pm17)$\,\kms{}. 
Here, observed velocities are given with respect to the Local Group centroid.

According to \citet{mei2007}, the Virgo cluster center,
distant from the Local Group at 16.5\,Mpc, has the mean velocity of
$V_{LG} = 1034\pm60$\,\kms{}. It corresponds to the infall velocity of the Local Group
towards the Virgo of $170\pm60$\,\kms{}, then 
the considered galaxy flock falls to the cluster with the velocity
of $\sim330$\,\kms{}. Being at a distance of about 11.8\,Mpc from the Virgo center, the
galaxies locate outside the zero-velocity radius of the Virgo, $R_0 = 7.2\pm0.7$ Mpc
\citep{kar2014}. Given their distance from Virgo and peculiar velocity, together with
similar data on other targets, one can calculate the Virgo cluster mass, assuming
spherical symmetry, with the formulae in \citet{tully_sh84}.

As it was noted above, gas-rich multiple dwarf systems, like DDO\,161 + UGCA\,319, are
attractive targets for studies of their structure and kinematics using HI-observations
with interferometer.

\section*{Acknowledgements}
This work is based on observations made with the NASA/ESA Hubble
Space Telescope, program GO-14636, with data archive at the Space
Telescope Science Institute. STScI is operated by the Association of 
Universities for Research in Astronomy, Inc. under NASA
contract NAS 5-26555.
The work in Russia is supported by the Russian Science Foundation
grant 14--12--00965. The authors thank Daniel Pisano for re-estimating
HI parameters for UGCA\,319 from ATCA observations and John Cannon for 
providing VLA  results that clearly resolve DDO 161 and UGCA 319.


\bibliographystyle{mnras}
\bibliography{ddo161text}   

\bsp

\label{lastpage}

\end{document}